# ReLEAF: A Socio-Technical Framework Bridging Custodians and Researchers for Trustworthy Data Sharing

Hibiki Ito[1], Chia-Yu Hsu[2], Hiroaki Ogata[2]

**Abstract**

Growing volumes of educational real-world data (ERWD) are being collected across learning platforms and institutional systems. Sharing these data within the Learning Analytics community offers substantial research opportunities, yet access remains limited by ethical, regulatory and governance constraints. Prior work has primarily focused on anonymisation techniques, but little attention has been paid to operational design of practical ERWD sharing, particularly how data custodians and researchers interact through privacy-preserving access mechanisms. To address this gap, we propose ReLEAF, a socio-technical framework that bridges data custodians and researchers by operationalising *two-stage data sharing*: 1) Differentially private synthetic data are shared for exploratory analysis, and 2) controlled real-data validation is performed on demand. Following a design-science research approach, we refine and formatively evaluate ReLEAF through three cycles involving 4 graduate students, 6 researchers, and 90 undergraduate students, respectively. Three design principles emerged through the cycles: **P1**) position privacy-preserving access mechanisms within the research workflow, **P2**) make the conditions for acceptable secondary use explicit and actionable, and **P3**) promote engagement with governance requirements rather than automate compliance decisions. Together, ReLEAF provides a concrete framework for trustworthy ERWD sharing, while the design principles offer transferable guidance for other data-sharing contexts.

## 1. Introduction

As artificial intelligence technology accelerates the expansion of the research and practice of Learning Analytics (LA), growing volumes of educational real-world data (ERWD) are being collected across large-scale digital infrastructures, such as Learning and Evidence Analytics Framework (LEAF; Ogata et al., 2022), the National Digital Education Architecture of India (DSEL, 2021) and the Taiwan Adaptive Learning Platform (MOE, 2025). By complementing experimental data, ERWD provides significant opportunities for advancing research to understand learning and empower learners and practitioners (Kuromiya et al., 2023; Mahajan, 2015; Okumura et al., 2026). Sharing these data within the LA community could substantially expand the range and scale of empirical inquiry in LA, yet privacy risks and governance requirements continue to constrain data sharing (Fischer et al., 2020; Joksimović et al., 2022).

Prior work has largely addressed the data access problem through anonymisation techniques, including $k$-anonymity (e.g. Stinar et al., 2024), differential privacy (e.g. Liu et al., 2025), and synthetic data generation (e.g. Ramos et al., 2026). Non-technological aspects, such as ethical frameworks (Drachsler & Greller, 2016; Pardo & Siemens, 2014; Slade & Prinsloo, 2013) and privacy perceptions of stakeholders (Longin et al., 2025; Mutimukwe et al., 2022; Salta et al., 2026), have also been studied. However, considerably less attention has been given to operational design of practical data sharing. Anonymisation alone does not enable meaningful data access; socio-technical arrangements between data custodians and researchers are also needed to operationalise trustworthy data sharing in practice. The MOOC Replication Framework (MORF) represented an important early attempt to operationalise controlled access to ERWD, but its restricted analytical environment limited usability and broader uptake within the research community (Baker & Hutt, 2025).

In response, this paper addresses the socio-technical problem in operationalising ERWD sharing. We particularly focus on data sharing with *semi-trusted researchers*. While semi-trusted researchers do not directly communicate with data subjects or data providers unlike *trusted* ones, their identities are known unlike *untrusted* third parties. That is, we are not aiming for a traditional trusted research environment (TRE), where researchers gain trust through months-long screening processes (Macnair et al., 2021), or publicly releasing open datasets that arbitrary untrusted parties can download. Instead, the semi-trusted access model should allow researchers to promptly start exploring data, while protecting privacy by design (Cavoukian, 2009).

To this end, we propose ReLEAF, a socio-technical framework of trustworthy secondary use of ERWD. It implements the recently proposed *two-stage* data sharing approach (Ito et al., 2026c): 1) Differentially private synthetic data are shared for exploratory analysis, and 2) controlled real-data validation is performed on requests. It is designed to balance utility and privacy for small-sample, high-dimensional ERWD (Ito et al., 2026c). Based on the approach of design-science research (DSR; Hevner et al., 2004), we iterate through three cycles of development and formative evaluation of the ReLEAF framework, built on the LEAF infrastructure. Throughout, the main research question of this paper is:

- **Main RQ**: How can semi-trusted secondary use of ERWD be operationalised through a socio-technical framework?

[1] Graduate School of Informatics, Kyoto University. Email: hibiki.itoo@gmail.com
[2] Academic Center for Computing and Media Studies, Kyoto University.

## 2. Literature review

### 2.1. Data sharing with semi-trusted researchers

Traditionally, access to sensitive data such as genomic and census records is provided in TREs (*trusted access*). Access methods vary from physical visits to secure remote access (Weise & Rauber, 2024). Researchers typically submit an application describing the purpose of data access and analysis plan, which is manually reviewed by the TRE staff (Cheah et al., 2017). After approval, researchers have access to pseudonymised data without any perturbative anonymisation measures. While this approach maximises research utility, the following three critical problems remain in the education domain.

First, months-long application review processes slow down the research process. For example, Macnair et al. (2021) report access times of 8–15 months across three UK health-data providers. Researchers in a cross-European study could expect approximately six months for data-permit approval and another six months for data delivery, with substantial variation across settings (Forster et al., 2025). Such long data access processes are poorly suited for ERWD because these data are routinely collected and shared (e.g. annually). Second, specifying a research purpose and an analysis plan before seeing data can be challenging. Because educational activities are highly local and ERWD are not collected for secondary studies, meaningful inspection of their variables and characteristics may be necessary before researchers can formulate a sensible analysis plan (van den Akker et al., 2025). Third, TREs rely too heavily on researchers' trustworthiness after access is granted. Although governance procedures establish trust before access, they cannot fully prevent or detect misuse or re-identification. Recent evidence of substantial re-identification risk in ERWD (Ito et al., 2025) and the closure of the inBloom project (Bulger et al., 2017) illustrate how such failures can erode stakeholder trust. Damaged trust can also affect self-disclosure behaviours of data subjects, and the validity of secondary analysis can in turn be degraded (Ito et al., 2026b).

On the other hand, an alternative to the TRE approach is to publicly release open datasets (*untrusted access*). This maximises the accessibility, solving the first two problems above: researchers do not need to wait for the access approval, nor specify the usage and analysis plan beforehand. However, the privacy risks cannot be mitigated as a free lunch. Sharing open datasets with untrusted, potentially adversarial third parties require stronger privacy protection algorithms, which necessarily trade-off the downstream utility. For example, Ito et al. (2026c) applied differentially private synthetic data generation (DP-SDG), arguably a gold-standard anonymisation technique, to ERWD, but only 36% of downstream analyses by researchers were successfully replicated on the real data. If we wish to preserve data utility, privacy must be compromised as a trade-off. Indeed ERWD have rarely been published as open data (Haim et al., 2023; Švábenský et al., 2026).

A *semi-trusted access* model is a hybrid of trusted and untrusted models, combining enhanced accessibility and heightened utility. Under the semi-trusted access model, researcher identities are known to a data custodian, but research purposes or plans are not meticulously reviewed before initial access. MORF represents an early attempt in this direction by offering a secure code execution environment for fine-grained MOOC data (Gardner et al., 2018). By limiting the execution output to common prediction metrics such as accuracy and precision, MORF protected individual privacy while avoiding traditional application processes (Hutt et al., 2022). While non-perturbative outputs establish the validity, users struggle to plan analysis and handle errors without seeing the data (Baker & Hutt, 2025). Table 1 compares the access models.

**Table 1.** Comparison of access models

| Access model | Where trust is placed | Accessibility | | Utility |
|---|---|---|---|---|
| | | Initial access time | Need to specify purpose / analysis plan beforehand | Downstream validity |
| Trusted (TREs) | Researchers | Long (often months) | Yes | High |
| Untrusted (open data) | Algorithms | Immediate | No | Low-moderate |
| Semi-trusted (MORF) | System restrictions | Short | Yes | High |
| **Semi-trusted (ReLEAF)** | **Algorithms + output review** | **Short** | **No** | **High** |

### 2.2. Two-stage data sharing

There are several possible data sharing methods under the semi-trusted access model (Carvalho et al., 2023). Traditional anonymisation methods, such as $k$-anonymity (Samarati & Sweeney, 1998) and $l$-diversity (Machanavajjhala et al., 2006), fail to robustly protect private information (Cohen, 2022). Synthetic data generation (SDG) is a modern approach, where a generative model is trained on real data, and then the model produces synthetic data to be shared. Nonetheless, SDG can be unpredictably vulnerable to privacy attacks, such as re-construction attacks (Dwork & Roth, 2014) and membership inference attacks (Stadler et al., 2022). Differential privacy addresses the problem of unpredictable privacy risks by offering mathematical privacy guarantee (Dwork et al., 2006), and DP-SDG is thereby often regarded as a gold-standard approach and has been adopted to untrusted access models (Abowd, 2018; Aktay et al., 2020; Hod & Canetti, 2025). Yet, synthetic data suffer from false discoveries (Ito et al., 2026c; Montoya Perez et al., 2024). This trade-off between privacy and downstream utility becomes even worse for ERWD, which is typically small in sample size and high in dimensionality (Ito et al., 2026c).

The two-stage data sharing approach was proposed in order to balance this tension, especially for ERWD (Ito et al., 2026c). As shown in Figure 1, Stage 1 is a typical data sharing process with DP-SDG. To complement the downstream validity issue of DP-SDG, Stage 2 involves a controlled real-data validation, which is often conducted under the trusted access model. Although the non-DP outputs of Stage 2 breaks the DP guarantee for the entire two-stage process, Ito et al. (2026c) demonstrate that the privacy risk of the entire process can be estimated through the framework of Gaussian membership inference privacy (Ito, 2024). This is contrasted with traditional non-DP anonymisation techniques that cannot robustly estimate privacy risks.

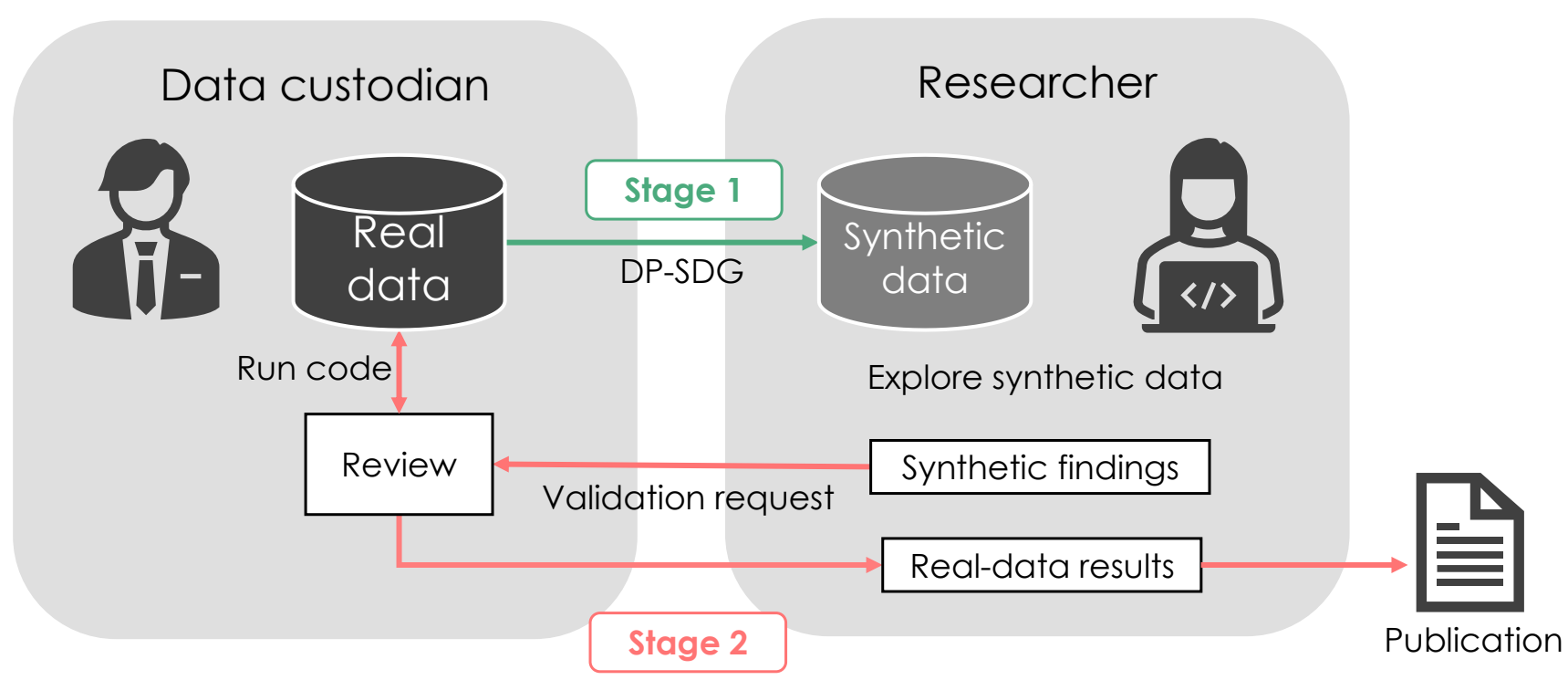

**Figure 1.** Overview of two-stage data sharing

Related work of two-stage sharing includes OpenSAFELY, a platform for the secondary use of electronic health records in the UK (Nab et al., 2024). OpenSAFELY provides researchers with dummy data for developing analysis scripts and allows subsequent real-data validation following a review process. However, because the dummy data are not generated from real data, they do not reflect the underlying data distribution. While such dummy data can facilitate script development, they may be less useful in educational research than in medical research. Medical researchers often have prior knowledge, expectations, or hypotheses regarding the underlying data distribution, whereas educational researchers may lack such information due to the localised and context-dependent nature of educational activities.

## 3. Literature-based initial design

The initial design of ReLEAF is proposed based on the literature (Figure 2). The framework consists of four components: **C1**) a structured organisation of datasites, datasets, assets, projects and validation requests, **C2**) implementation of the two-stage sharing, **C3**) publicness-first review criteria and **C4**) an LLM-to-human review process.

The concepts and organisation of datasites, datasets, assets, projects and requests are borrowed from the PySyft framework[3], as adopting a generic open-source framework makes implementation easier. While PySyft expects a real-mock pair of CSV files as an asset, we use a pair of real (i.e. pseudonymised) and DP synthetic CSV files, where the synthetic data is generated from the corresponding real data with DP. The pseudonymised and DP synthetic data are obtained from the LEAF infrastructure.

The two-stage sharing method is implemented through two UIs: Lab and Review. On Lab researchers download synthetic datasets from distributed datasites and then use them locally. If they wish to validate their findings on synthetic data for e.g. publishing a paper, they may create a project describing the research objective and the public interest rationale. While a project specifies the scope and publicness, each validation request in a project contains a more specific research question and associated Python code. On Review data custodians review each request manually, remotely run code on datasites and approve requests so researchers can download real-data outputs on Lab.

We define a concise set of review criteria composed of technical requirements and a publicness requirement. The technical requirements are due to the adoption of PySyft. Usually in TREs, real-data outputs are carefully reviewed for their disclosure risks, and researchers are required to apply statistical disclosure controls, such as small-cell suppression (Griffiths et al., 2024). However, ReLEAF is structurally different from such TREs in that only necessary validation of synthetic findings is performed on ReLEAF, while every finding requires output checking on TREs. Because synthetic data do not contain real individual records, findings on synthetic data are necessarily group-level trends. Thus, the publication-ready criterion is intended to discourage individual-level, potentially disclosive outputs without imposing detailed statistical disclosure control requirements. This approach is also expected to help researchers comprehend the criteria and reduce the cost of training to become "safe researchers" (Desai et al., 2016; Macdonald et al., 2023).

[3] https://github.com/OpenMined/PySyft

The LLM-to-human review process supports researchers in preparing high-quality requests, while preserving human accountability. Three LLMs, prompted to review a request based on the above proposed criteria, independently judge a request and give a comment. Based on the LLM review results, a researcher either submits the request for human review or withdraws it to create a revised one. This process is intended to reduce manual review time by identifying trivial issues regarding the review criteria (Baker & Hutt, 2025).

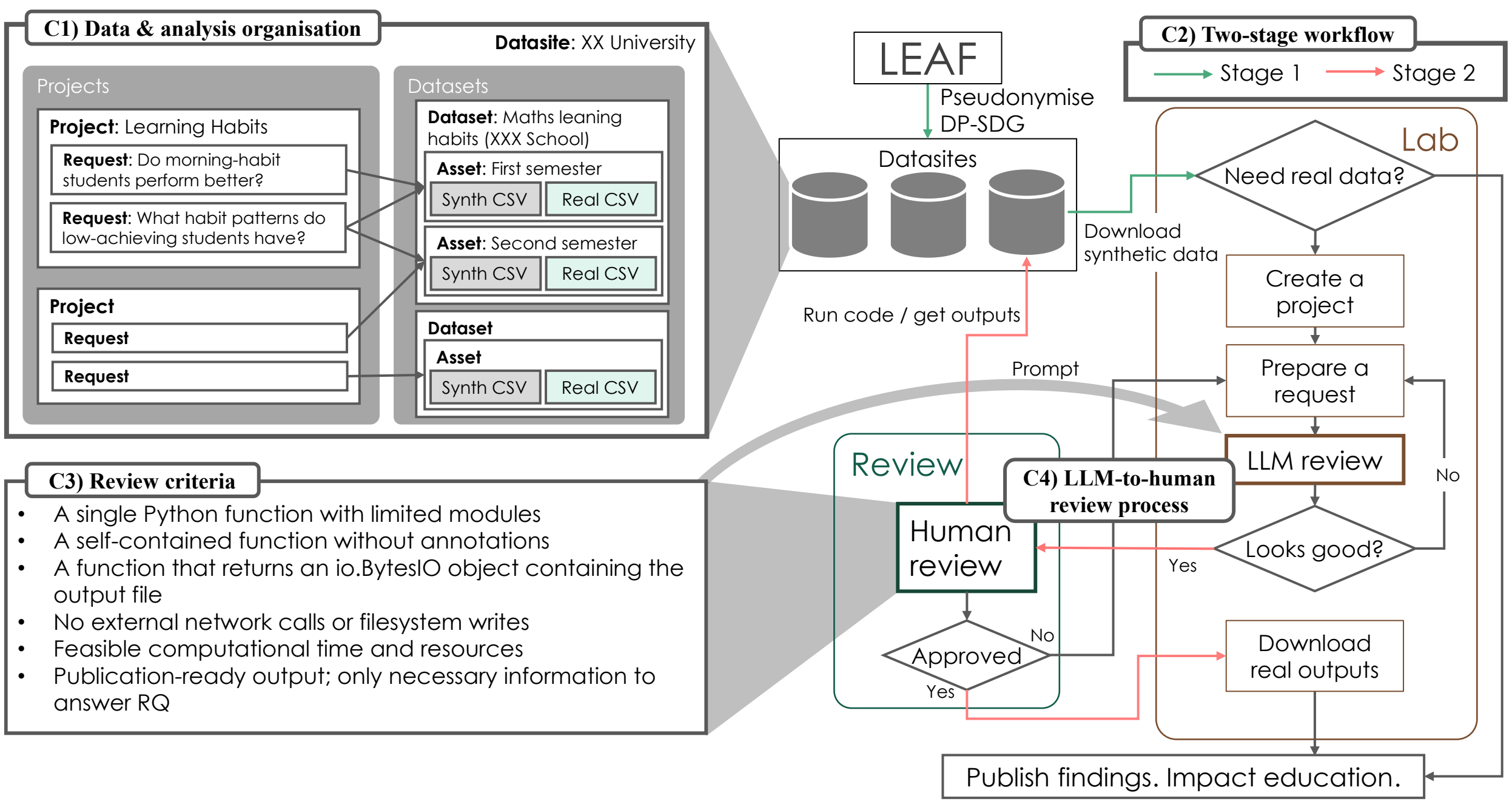

**Figure 2.** Literature-based initial design of ReLEAF

## 4. Iterative development and formative evaluation

We take the approach of DSR because the objective of this study is to develop and refine a socio-technical artefact for a practical data-sharing problem, rather than only to evaluate the proposed framework (Hevner et al., 2004; Peffers et al., 2007). The initial design of ReLEAF is iteratively refined and formatively evaluated across three cycles. Throughout, informed consent was obtained from the data subjects, including LEAF users and study participants. One of the authors, an expert in statistical disclosure control, reviewed each validation request. Since the system is still a prototype and cross-border data sharing is legally restricted in Japan, all study participants are affiliated with domestic universities or research institutions, and they are requested to delete downloaded synthetic data after the studies. The study was approved by the institutional review board of Kyoto University (No. 2023-001). Detailed study materials, including participant instructions, questionnaires, AI prompts and dataset descriptions, are available in the Supplementary material.[4]

### 4.1. Cycle 1: Initial user study

Since real-data validation introduces additional information-theoretic privacy loss and operational costs associated with output checking, it is important to understand to what extent synthetic data alone (Stage 1) can support secondary use and when real-data validation (Stage 2) becomes necessary. Additionally, the comprehensibility of the proposed publicness-first review criteria remains unknown. Thus, we set the following research questions for the initial user study:

- **RQ1-1**: To what extent can users perform secondary use with synthetic data alone and when does real-data validation become necessary?

[4] https://anonymous.4open.science/r/releaf-paper-supplementary

- **RQ1-2**: To what extent do users develop an understanding of ReLEAF's review criteria through sustained use of the system?

#### 4.1.1. Methods

A formative user study was conducted with four graduate students over five weeks. Three participants majored in educational technology and one in another field of computer science. In the first week, participants went through a tutorial of ReLEAF and initial exercises of output checking. Then, we created the common project "Learning Habits and Outcomes" on which everyone worked independently. The project objective was set to the following:

- **Project objective**: Understanding the relationship between learning habits and outcomes and identifying potential issues with learners that would need support.

Three LEAF datasets from years 2022 to 2024 were prepared. Each dataset contains seventh graders' ($N = 120$ per year) learning outcomes (low, mideum and high) and time spent on BookRoll over 17 weeks, an e-book system within LEAF (Ogata et al., 2017). The BookRoll session data was aggregated by week and time windows—morning, afternoon, evening and overnight—following the method of Hsu et al. (2023). We designed a multi-shot synthesis setting, where similar data of different learner cohorts arrive regularly (Ito et al., 2026a):

1. Dataset 2022 was uploaded to ReLEAF, and students analysed it during the second week.
2. Dataset 2023 was uploaded at the beginning of the third week, and students analysed both datasets during the week.
3. In the fourth week, students shared findings during the class.
4. Finally, dataset 2024 was uploaded at the beginning of the fifth week, and students analysed it as well as the previous datasets during the final week.

This way, we aimed to emulate a three-year secondary-use project. We generated DP synthetic data by following the LLM-based multi-shot synthesis method of Ito et al. (2026c) with privacy parameters $\varepsilon = 1$ and $\delta = 10^{-3}$. Specifically, summary statistics were calculated from real data with DP, and then LLMs generated Python scripts to produce synthetic data that align with the given context, schema and the DP statistics. The generation of the second and third datasets was informed by real-data validation results of previous years so that synthetic data quality is expected to cyclically improve (Ito et al., 2026c, 2026a).

A semi-structured focus group was conducted at the end of the study because it generates rich *language in vivo* which helps design subsequent studies, thereby suitable for a formative study (Tracy, 2012). A moderator prompted the below questions, followed by free discussions:

- To what extent did you achieve the project objective with synthetic data alone? (RQ1-1)
- To what extent did you achieve the project objective with the addition of real-data validation? (RQ1-1)
- To what extent do you understand the review criteria? (RQ1-2)

For analysing the focus group data, we used the Steps for Coding And Theorising (SCAT), as it is suitable for one-time, small-scale data (Otani, 2008). SCAT codes utterances into themes and constructs and then contextualises them back into a single storyline. Given evolving evidence that LLMs can effectively support qualitative analysis (Hayes, 2025), a few-shot-LLM-based SCAT pipeline was developed.[5] Specifically, a human researcher, GPT-5.4, Gemini 3.1 Pro and Claude Opus 4.6 generated themes and constructs independently. The researcher filters LLMs' outputs that are important but missing in the human output, and a storyline is then manually created from the updated human output.

#### 4.1.2. Results

We first report the number of requests over the study period in Figure 3. Period 1 started when dataset 2022 was uploaded (week 2), period 2 started when dataset 2023 was uploaded (week 3 and 4), and period 3 started when dataset 2024 was uploaded (week 5). 38 requests were created in total, 4 requests were rejected, and 8 requests were withdrawn after LLM reviews. For RQ1-1, the number of requests increased over time, suggesting that participants increasingly utilised real-data validation as the study progressed. For RQ1-2, we observe that the approval rate did not increase over time, providing no clear evidence of improved adherence to the review criteria over time.

[5] https://github.com/hibiki-i/scat-llm-pipeline-v2

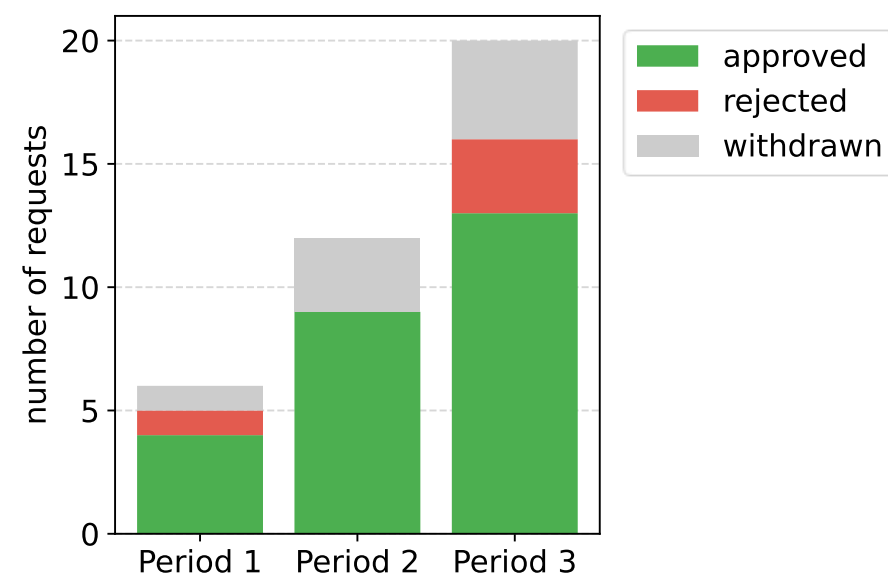

**Figure 3.** Number of requests over time during the study

For RQ1-1, Figure 4 shows the storyline derived by SCAT on the focus group data. Underlines indicate themes and constructs derived from participants' utterances through SCAT, where bolded ones are extracted by LLMs. The result suggests that synthetic findings are perceived as infeasible to publish or apply to educational practice. The participants reported that real-data validation becomes necessary for such downstream integration, unless the fidelity of synthetic data is guaranteed. Discrepancies between synthetic and real data could damage credibility in synthetic data, and transparency in the SDG process shapes the analysis plan. As a trade-off with privacy protection, the review process can constrain the speed of research processes.

The participants observed that LLM-generated synthetic data tended to produce hypothesis-confirming results that are not validated on real data, pointing out the infeasibility to achieve synthetic-data-alone scientific discoveries and practice applications. This included particular concerns about longitudinal risk of false-discovery accumulation. To mitigate false discoveries, participants demanded SDG transparency, which **shapes the analysis plan**. For conditions when real-data validation becomes necessary, a participant optimistically viewed future synthetic-data-alone analysis given sufficient fidelity, while still showing **apprehension towards practical and academic applications without real-data validation**. On the other hand, another participant posited **delegation of validity assurance to data providers**, demanding either validity guarantee or real-data validation. Although real-data validation enables scientific publications and practice applications, a participant experienced damaged credibility of synthetic data with real-data invalidation and the constraints on the speed of the research process due to manual reviews.

**Figure 4.** Storyline of the focus group for RQ1-1

The SCAT storyline for RQ1-2 is shown in Figure 5. The participants reported general understanding of the review criteria and the first-time comprehensibility of technical requirements. The initial exercises of output checking potentially help users grasp the idea of publication-readiness. Yet, assessing disclosure risks appears challenging. Although disclosure risks are not explicitly mentioned in the review criteria, some requests were rejected due to unnecessarily disclosive (i.e. too detailed) outputs, which were judged as not ready for publication. In addition, the result suggests that the black-box one-off LLM reviews can introduce potential cognitive offloading such that users rely on heuristic strategies like majority voting, rather than deeply engage with the review criteria.

The participants reported general understanding of the review criteria after the study. Technical requirements were perceived as relatively easy to grasp at first glance, although one participant indicated a demand for clearer explanation of technical concepts such as `io.BytesIO`. In contrast, the initial exercises revealed **early misconceptions of disclosure risks** among some participants, and all participants reported low confidence in assessing such risks. Several participants noted that their initially optimistic self-evaluations did not align with the LLM reviews, while others raised concerns about the black-box nature of those reviews. Consequently, multiple participants reported relying on majority voting across LLM reviews, suggesting a degree of cognitive offloading to the AI-assisted review process.

**Figure 5.** Storyline of the focus group for RQ1-2

#### 4.1.3. Design implications

Although the initial user study indicates the necessity of the validation process for publishing and applying findings on ERWD, it revealed challenges in the human-to-LLM review process and the review criteria comprehensibility. While the LLM-supported review process potentially reduces review costs by identifying minor issues, it should enable users to actively engage with improving the request quality, rather than to overrely on it as a validation filter. This would include providing opportunities for users to ask for clarification of review decisions, addressing the black-box problem of LLM review. Additionally, since the results suggest the proposed publicness-first review criteria can confuse users about disclosure risks, explicit guidelines about such risks may be needed.

## 4.2. Cycle 2: Researcher field evaluation

Based on the initial user study, we refine the framework and evaluate it with real researchers, a more realistic sample of the user population. Due to the small sample size, the evaluation is exploratory and formative.

### 4.2.1. Refinement and research questions

We update two components: the LLM-to-human review process and the review criteria. First, the mandatory LLM review step is replaced with a voluntary chatbot interaction (Figure 6). The chatbot has two modes: Review and Dev. In the Review mode, the AI reviewer can review a request and answer follow-up questions to clarify the judgement. In the Dev mode, the AI reviewer can edit code to e.g. meet the technical requirements or draft a compliant code snippet. This update aims to address the black-box problem of LLM reviews and avoids majority-vote heuristics. Zero-shot GPT-5.4 was used for the AI reviewer.

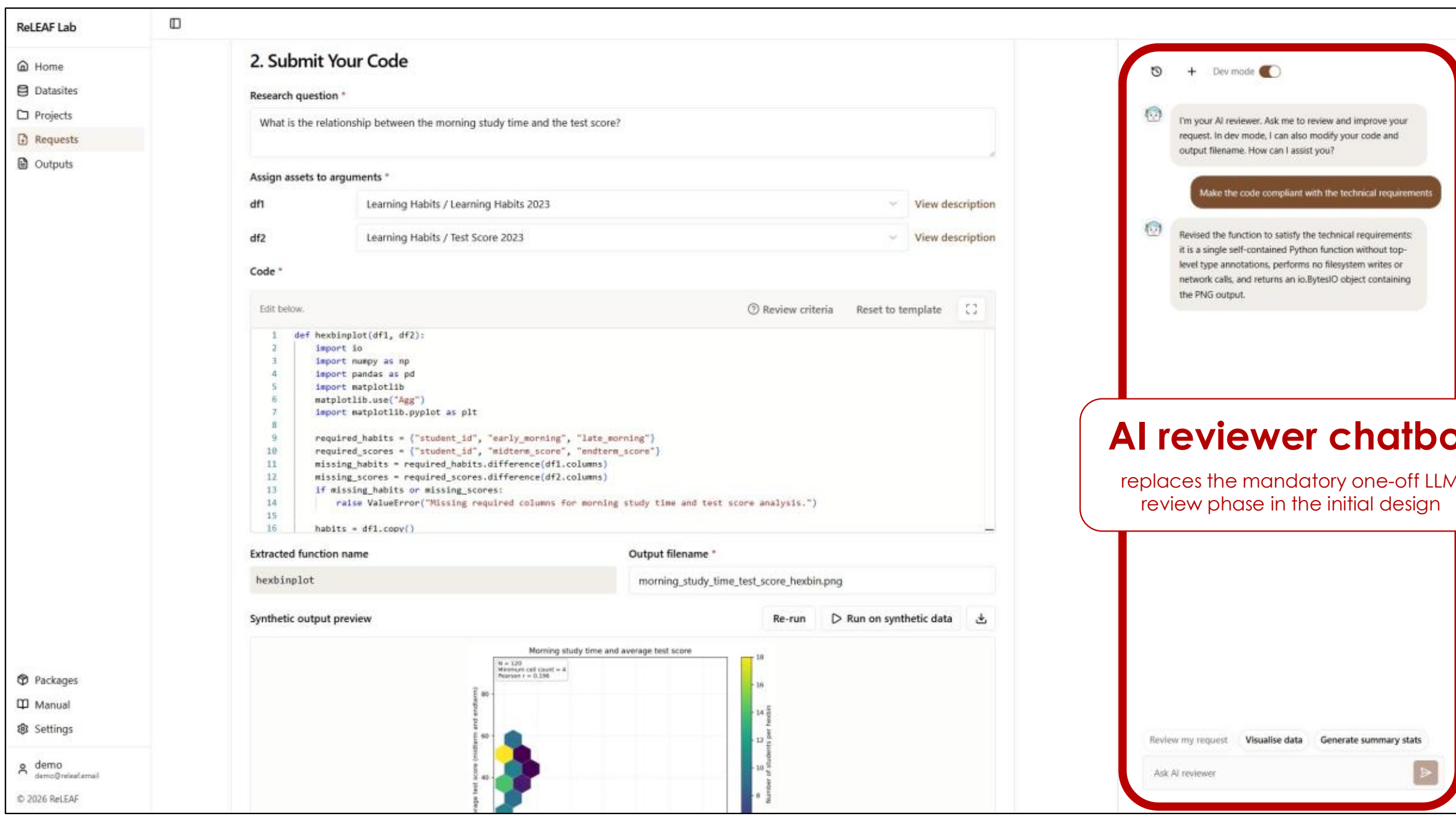

**Figure 6.** Screenshot of the request creation page for Cycle 2

Second, the disclosure risk assessment is explicitly added to the review criteria. The revised criteria are the following:

- A single Python function with limited modules
- A self-contained function without annotations.
- A function that returns an `io.BytesIO` containing your output file.
- No external network calls or filesystem writes.
- Fit resource limits: runtime ≤ 10 minutes, CPU/memory within 2 vCPU/8GB, and output ≤ 10MB. (The resource constraints depend on the deployed instance.)
- Publication-ready output; only necessary information to answer RQ.
- No individual-level information allowed.

To formatively evaluate these refinements, we set the following research questions for Cycle 2:

- **RQ2-1**: How do researchers use the AI reviewer chatbot during request preparation?
- **RQ2-2**: What challenges arise when researchers use the revised ReLEAF workflow and review criteria?

### 4.2.2. Methods

Having called domestic researchers for voluntary participation to a pilot, 17 researchers registered for the pilot by answering a pre-pilot survey, but only 6 of them actively used the system during the pilot period. The pre-pilot survey includes researcher background questions, three initial exercises of output checking like Cycle 1 and open-ended forms about the exercises. All three exercise questions deal with disclosure risks, and one of them also addresses publication-readiness (i.e. only necessary information to answer RQ). We use the pre-pilot survey responses of all 17 researchers, but we only count the 6 active users as the pilot participants. Table 2 describes their backgrounds.

**Table 2.** Backgrounds of pilot participants

| ID | Position | Major | Familiarity with DP | Familiarity with SDG | Number of correct exercises |
|---|---|---|---|---|---|
| P1 | PhD candidate | Educational Technology | Aware | Aware | 1 |
| P2 | Associate Professor | Education | Aware | Informally familiar | 2 |
| P3 | Master's student | Computer Science | Aware | Aware | 3 |
| P4 | Professor | Educational Technology | Aware | Aware | 0 |
| P5 | Professor | Educational Technology | None | None | 2 |
| P6 | Researcher | Computer Science | Aware | Aware | 2 |

We shared a learning habits dataset with P1 to P5 and a self-directed learning dataset with P6, because data sharing with P6 was part of research collaboration. The former dataset consists of similar time series as the dataset for Cycle 1 and mid-term and end-term test scores: seventh graders' time spent on studying mathematics on BookRoll was aggregated for each of 18 weeks and for each of six time-of-day windows. The dataset contains four cohorts: 2022 (N=119), 2023 (N=120), 2024 (N=118) and 2025 (N=117). The latter dataset contains xAPI logs of seventh graders with BookRoll and the GOAL system, a self-directed learning-support system, in a mathematics course over approximately two months. For both datasets, we used the LLM-based DP-SDG with $\varepsilon = 1$ and $\delta = 10^{-3}$ as for Cycle 1.

The pilot spanned over 10 weeks, and the participants were required to download synthetic data and submit at least three validation requests, so all participants experience the ReLEAF workflow. A post-pilot survey used a five-point Likert scale to assess the comprehensibility of the two-stage data sharing method, acceptability of ReLEAF as a secondary-use framework, perceived usefulness of synthetic data for developing analysis and perceived understanding of review criteria. It also contained open-ended feedback forms about AI reviewer chatbot and the balance of privacy protection and research utility on ReLEAF.

For RQ2-1, thematic analysis was performed on the chat logs to explore usage patterns (Braun & Clarke, 2006; Howard & Hoffman, 2018). For RQ2-2, rejected requests were qualitatively inspected as case studies, and survey responses were reviewed and descriptively summarised. Due to the small scale of the pilot, results should be seen as exploratory.

#### 4.2.3. Results

We received 40 requests in total, of which 29 requests were approved. For RQ2-1, each message sent by a participant to the AI reviewer received one code. The number of messages per request and the frequency of each code are shown in Figure 7 for descriptive purposes. Around half of the requests were submitted without interactions with the chatbot.

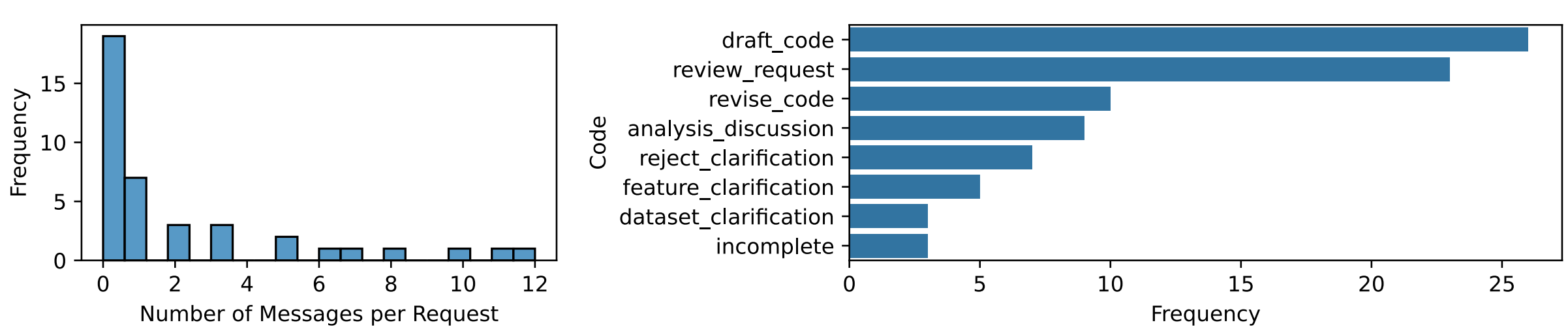

**Figure 7.** Distribution of the number of messages per request (left) and the code frequencies (right)

First, the participants used the AI reviewer for *discussing request development*, which is essentially what the refinement from Cycle 1 aims to support. Several participants also reported in the post-pilot survey that the AI reviewer was helpful with the final check of technical and privacy-protection requirements. A case of P3 represents the potential and challenges of this usage: P3 attempted to perform linear discriminant analysis outputting the top five discriminating weeks and their separation accuracy and AUC for each year. However, the AI reviewer rejected this analysis due to the disclosure risk of small-cell outputs, and P3 had several turns of discussing the risk and revising the code. The AI reviewer consistently rejected the request, but P3 finally submitted it for human review. Then the request was eventually approved by the human expert. Presumably based on this experience, P3 stated in the post-pilot survey that the AI reviewer appears not so accurate and sometimes too confident about their judgements.

Second, some participants utilised the AI reviewer for *analysis development* (`draft_code`, `revise_code`, `analysis_discussion`). For example, P5 asked for drafting code that cross-tabulates *morning-* and *night-habit* students (`draft_code`), and the AI reviewer edited code to output a table of statistics while requesting clarification of definitions of morning and night habits. Then P5 asked the AI reviewer to classify students into three categories by morning-study time (`revise_code`). Although the Dev mode was initially designed to help researchers develop ReLEAF-compliant code, this type of usage suggests that the AI reviewer can support researchers not only in the ReLEAF-specific request-creation process but also in analysis development in general.

Regarding RQ2-2, in the pre-pilot survey, the overall correctness of the initial exercises was 47%, indicating the difficulty of conveying the review criteria without concrete examples. Several participants reported the exercises revealed initial misconceptions of the review criteria. Additionally, multiple participants raised a concern that the criterion "only necessary information to answer RQ" may be difficult to consistently apply. In the pilot, 11 requests were rejected, and their reasons can be categorised into two types: (1) disclosure risks and (2) missing project details. Some rejected requests contained MIN/MAX queries or a scatter plot, despite the warning in the initial exercises. This suggests the potential need for a more refined tutorial or more intuitive user experience, rather than the criterion itself. Other requests had project description filled with placeholders (e.g. "test"), perhaps because users tested project creation and forgot to update the description. While the project validity is not explicitly mentioned in the review criteria, we explained prior to the pilot that project descriptions are also part of review with regard to publication-readiness. Nonetheless, in the post-pilot survey, the participants agreed that the criteria are mostly comprehensive.

In the pre-pilot survey, multiple participants indicated that the unfamiliar two-stage workflow can be confusing, especially the role of synthetic data, which suggests the need to clarify the workflow from the user perspective. Nonetheless, in the post-pilot survey, the participants positively responded about the comprehensibility of the two-stage data sharing method and acceptability of ReLEAF as a secondary-use framework. P3 pointed out that the discrepancy between real and synthetic data would raise the need to validate every hypothesis, potentially slowing down the research process. Yet, P4 showed an understanding that the discrepancy is acceptable to some extent because sharing otherwise inaccessible data is beneficial to the research community.

#### 4.2.4. Design implications

Although the results are exploratory, main design implications are reframing of the roles of synthetic data and validation requests and revision of review criteria. While the two-stage data sharing is a technical solution, a socio-technical framework should position each stage in the research workflow, so users can adopt the framework in their secondary analysis of ERWD. Additionally, the requirement of only necessary information in request output may be too strong and can be more softly framed. Disclosure risks and project relevance can also be more explicitly mentioned in the review criteria. As the generic AI review can sometimes be inaccurate or overconfident, more uncertainty could be introduced. In other words, the AI reviewer should focus on trivial issues to reduce the cost of manual review, while helping users engage with creating a trustworthy validation request.

### 4.3. Cycle 3: Operationalising at scale

Finally, we further refine the framework and aim to formatively evaluate it at a larger scale. Since employing a large sample of active researchers is impractical, Cycle 3 involves undergraduate students as proxy novice users.

#### 4.3.1. Refinement and research questions

Three components are refined. First, we reframe the two-stage workflow as *hypothesis generation* (Stage 1) and *validation* (Stage 2), explicitly communicating these roles in the tutorial.

Second, the review criteria are substantially revised: the requirements are reorganised into three categories; the "necessary information" criterion is reframed softly; and the prohibition of MIN/MAX queries, the requirement of project relevance and a requirement of rounding numeric outputs are explicitly added. While traditional TREs require more advanced disclosure control measures, such as small-cell suppression and subsampling for regression analysis (Griffiths et al., 2024), we do not impose these requirements in this prototype. Instead, we adopt simpler requirements intended to reduce disclosure risks while limiting the researcher training burden (Desai et al., 2016). The new review criteria are shown in Table 3.

**Table 3.** Revised review criteria for Cycle 3

| Category | Requirement |
|---|---|
| Technical | Submit exactly one Python function. |
| | The function must be self-contained and use only permitted standard/scientific modules. |
| | Do not use type annotations. |
| | The function must return an `io.BytesIO` object containing the output file. |
| | Do not make external network calls. |
| | Do not write to the filesystem. |
| | Stay within resource limits: runtime ≤ 10 minutes, memory/CPU within 2 vCPU / 8 GB, and output file ≤ 10 MB. |
| Disclosure | Do not output individual-level records or values. |
| | Do not use MIN or MAX queries. |
| | Do not create scatter plots or other plots where each mark corresponds to one individual. |
| | Round numeric outputs to reasonable precision, e.g. 3 significant figures. |
| Validity | Output must be publication-ready. |
| | Include only information that directly contributes to answering the research question; exclude tangential or redundant information. |
| | The analysis must be relevant to the project description. |

Third, we accordingly prompt the AI reviewer to review a request by category, aiming for better transparency. In addition, instead of approval-or-rejection judgements, the AI reviewer is instructed to output either approval, rejection or *unsure*. This aims to address the overconfidence issue observed in Cycle 2 and to potentially mitigate users' overreliance on AI reviews observed in Cycle 1 (Kim et al., 2024).

Note that the participant sample is not representative of the researcher population. We thereby set the following research question for this cycle:

- **RQ3-1**: How do novice users enact the revised hypothesis-generation-to-validation workflow?
- **RQ3-2**: What types of issues arise in validation requests under the revised review criteria?
- **RQ3-3**: What operational burden does human review impose under the revised framework?

#### 4.3.2. Methods

A larger-scale study was conducted with 90 undergraduate students in an introductory data science course of teacher education. The same learning habits dataset as in Cycle 2 was shared. The students have basic familiarity with data science, but programming experience of Python is not guaranteed. Thus, the participants initially had a request-creation practice, and these practice requests were manually reviewed. The participants checked the validation results and gave feedback on how synthetic-data results differ from real-data results, so that all participants experienced the hypothesis-generation-to-validation workflow on ReLEAF. Next, the participants were given two tasks: creating a project with a given objective and publicness rationale and submitting a request under that project with one research question chosen from three options (Task 1) and creating an original project and submitting an original request under that project (Task 2). The submitted requests were manually reviewed, and the participants were requested to give feedback on the difference between synthetic- and real-data results.

Each request was reviewed according to the revised criteria. Due to time constraints, we set an additional review policy that code is not reviewed in detail when the output is interpretable from the labels and the research question. Note that the purpose of code review is to check the consistency of output files. For example, an adversarial request can contain a valid research question and a valid-looking output file with intentionally incorrect figure labels to disguise disclosive information. Our review policy in this study assumes no such cases.

For RQ3-1, the following key events are identified from the usage logs: metadata page views, project creation, RQ edits, code edits, interactions with the AI reviewer, request submission, real-data validation (synthetic-real discrepancy feedback). The usage patterns of these key events are visualised with a directly-follows graph (DFG). For RQ3-2, rejected requests are qualitatively reviewed. For RQ3-3, review time is estimated based on the interaction logs of the reviewer for both Cycle 2 and 3. We explore the relationship between review time and several aspects of a request, including the output format, as prior work describes it as a factor contributing to review time (Fisher, 2023).

#### 4.3.3. Results

100 and 103 requests were submitted and reviewed for Task 1 and 2, respectively. Regarding RQ3-1, Figure 8 displays the DFG of user interactions. After viewing the metadata, users move on to the request creation phase, rarely coming back to the metadata page. In the request preparation phase, the natural path from the RQ edit, the code edit to the AI chatbot is emphasised, indicating a common workflow. Frequent self-transition of AI interactions highlights that participants iteratively developed and discussed the analysis with the AI reviewer. In addition, the reverse arrows from the AI chats to the RQ edit and code edit suggest that participants revised research questions and code through interactions with the chatbot. Most requests are submitted after the final check with AI. These patterns illustrate the central role of the AI reviewer in the hypothesis generation phase for novice users. Nonetheless, while 190 requests were approved in total, only 124 of them were explicitly checked for validation by users, suggesting limited engagement in hypothesis validation.

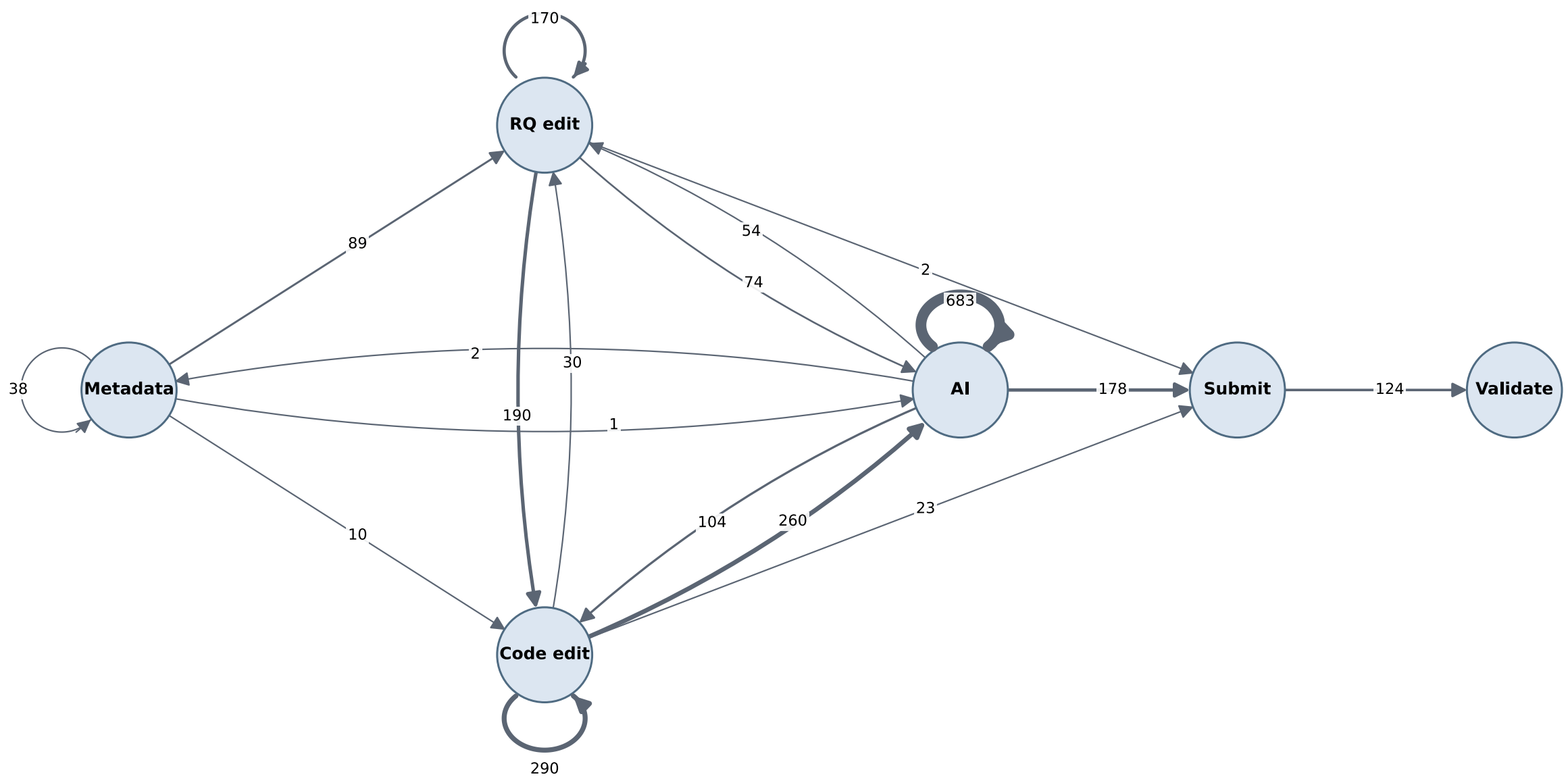

**Figure 8.** Directly-Follows Graph of user interactions with the system

Regarding RQ3-2, 3 out of 100 (3.0%) and 10 out of 103 (9.7%) requests were rejected. The higher approval rate may partly reflect the AI-assisted request preparation process, as the AI reviewer is prompted to write ReLEAF-compliant code. Qualitatively reviewing the rejected requests among the task, four types of issues are observed: (1) technical issues such as Python syntax errors and erroneous output file names, (2) insufficient formulation of research questions (e.g. only analysis titles), (3) misconceptions about available datasets and (4) potential disclosure risks. First, the technical issues are not essentially a failure of the revised framework, but rather an implementation issue, because they could be prevented by systematic validation before submission. The second and third issues would be largely attributed to participants' lack of experience in data science. Nonetheless, these issues could also be addressed with systematic support in future development.

Few requests were rejected due to disclosure risks, potentially reflecting the conservative characteristics of AI-assisted analysis development. A participant created a request for a simple correlational analysis between the proportion of active weeks (i.e. weeks with at least one minute of BookRoll activity) and the test scores. Table 4 shows the synthetic-data output for year 2023. The metrics for 0-25% and 25-50% coverage groups are suppressed to prevent individual-level information disclosure. Although this may seem to control disclosure risks well, the request was rejected because this output format can indicate that only one student belongs to the most inactive group, effectively revealing individual-level information. According to the AI reviewer chat logs, the participant developed the request with the help of AI, and the AI reviewer indeed approved the request before submission. Overall, although such a case would not be particular to novice users, assessing disclosure risks may be a challenge for both users and the zero-shot AI reviewer.

**Table 4.** Example output of a rejected request

| Coverage group | Number of students | Mean activity coverage | Mean mid-term | Mean end-term | Mean average |
|---|---|---|---|---|---|
| 0-25% | 1 | — | — | — | — |
| 26-50% | 0 | — | — | — | — |
| 51-75% | 10 | 0.656 | 47.600 | 44.200 | 45.900 |
| 76-100% | 109 | 0.982 | 44.174 | 41.138 | 42.656 |

For RQ3-3, the cumulative distributions of review time are shown for Cycle 2 and 3 in Figure 9. Most requests are reviewed within 30 minutes for Cycle 2 and 10 minutes for Cycle 3. Overall, the review time is greatly reduced for Cycle 3, partially due to our code-review policy.

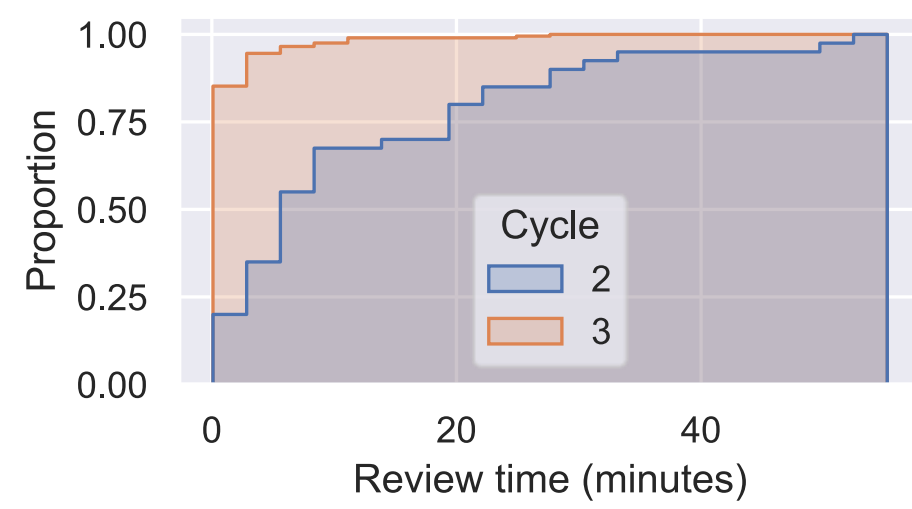

**Figure 9.** Cumulative distribution of review time for Cycle 2 and 3

Figure 10 shows the relationship between review time and the number of input assets, abstract syntax tree (AST) nodes, lines of code, cyclomatic complexity and the output format. Across the four scatter plots, the patterns are unclear, indicating that these metrics do not reflect the factors contributing to the operational cost. Regarding the output format, Cycle 3 exhibits a clearer pattern that table outputs take more time to review, yet this is likely attributed to our review policy, where table outputs often require code review to understand the output content. Since Cycle 2 participants are more experienced researchers, these results and the generally longer review time for Cycle 2 together suggest that semantic complexity of analysis may have a stronger link to the operational cost of review time.

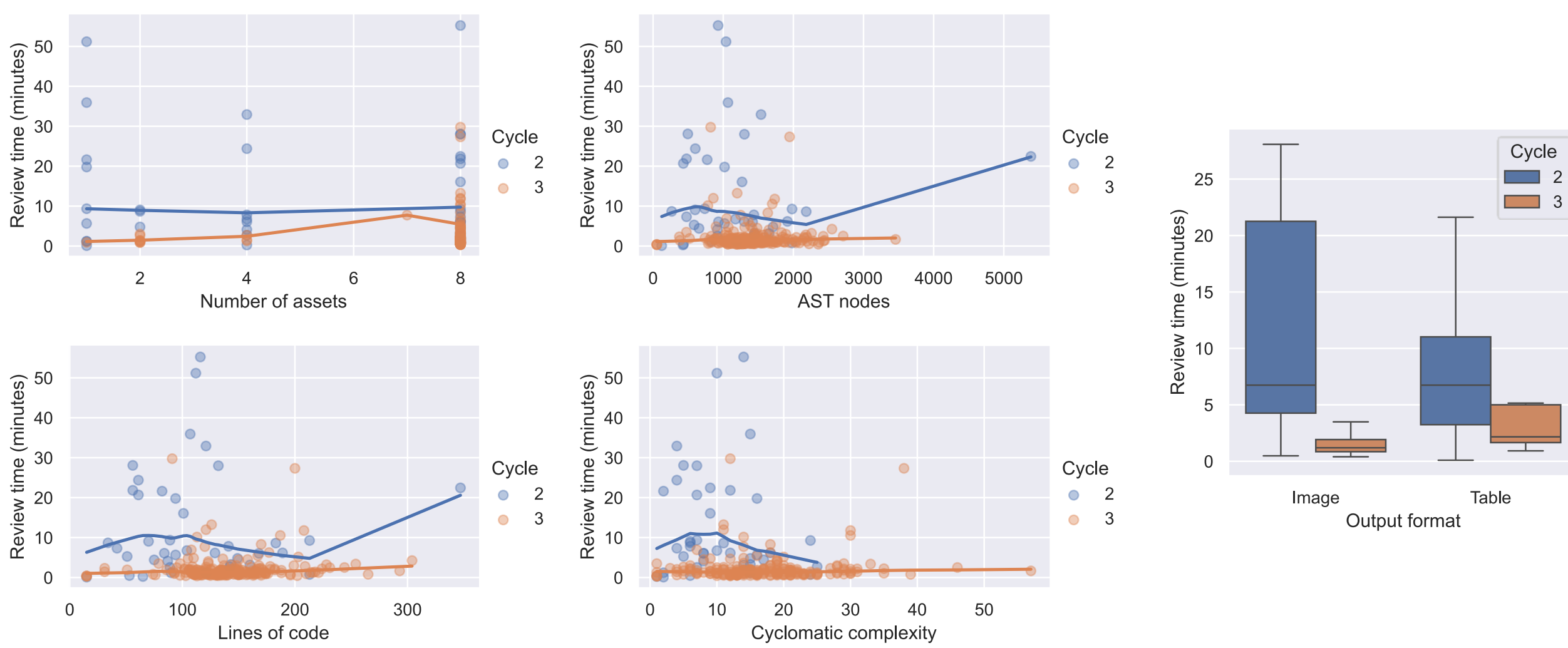

**Figure 10.** Relationship between review time and analysis complexity metrics for Cycle 2 and 3

#### 4.3.4. Design implications

The AI chatbot should remain assistive, rather than authoritative, enabling user engagement and human accountability. More technical refinement might be needed to prevent inaccurate reviews with overconfidence. Also, the three-part review structure appears operationally usable, as observed rejection reasons can be meaningfully captured in the three dimensions. Moreover, the review time analysis provides preliminary evidence about the operational burden of controlled validation, an important consideration for scalability. For wider adoption in the research community, operational burden of manual review should be part of the framework design.

## 5. Discussion

### 5.1. Answer to the main RQ

Through three DSR cycles, we refined the framework and derived three design principles (Figure 11). The empirically refined ReLEAF framework retains the original data and analysis organisation (**C1**), reframes two-stage sharing as hypothesis generation and validation (**C2**), organises review criteria into technical, disclosure and validity requirements (**C3**), and replaces one-off LLM reviews with AI-assisted request preparation under human oversight (**C4**). Together, ReLEAF integrates these key components into a socio-technical framework that bridges data custodians and researchers for trustworthy secondary use of ERWD.

Although ReLEAF is implemented on the LEAF infrastructure, its components are designed to be infrastructure-independent. To implement the framework, a data custodian needs to generate DP-synthetic datasets and upload a pair of real and synthetic files. Integration of other digital infrastructures in education is left for future work.

The derived design principles are as follows: **P1**) Position privacy-preserving access mechanisms within the research workflow. Cycle 1 revealed that participants perceived real-data validation as necessary for downstream application, while Cycle 2 identified confusion about the respective roles of synthetic data and validation requests. These findings motivated reframing the two-stage mechanism as hypothesis generation and validation, highlighting the importance of positioning privacy-preserving access mechanisms within researchers' workflows. **P2**) Make the conditions for acceptable secondary use explicit and actionable. Across the three cycles, participants encountered difficulties interpreting disclosure requirements, and rejected requests revealed persistent misconceptions despite successive revisions of the criteria. These observations suggest that simplifying governance requirements alone is insufficient; access conditions should be communicated explicitly and supported by actionable guidance, even at the expense of additional complexity. **P3**) Promote engagement with governance requirements rather than automate compliance decisions. Across the three cycles, AI-assisted review revealed risks of user overreliance, inaccurate judgements and overlooked disclosure risks. These findings highlight the limitations of automating compliance decisions, suggesting that AI should instead support researchers' engagement with governance requirements while preserving human accountability.

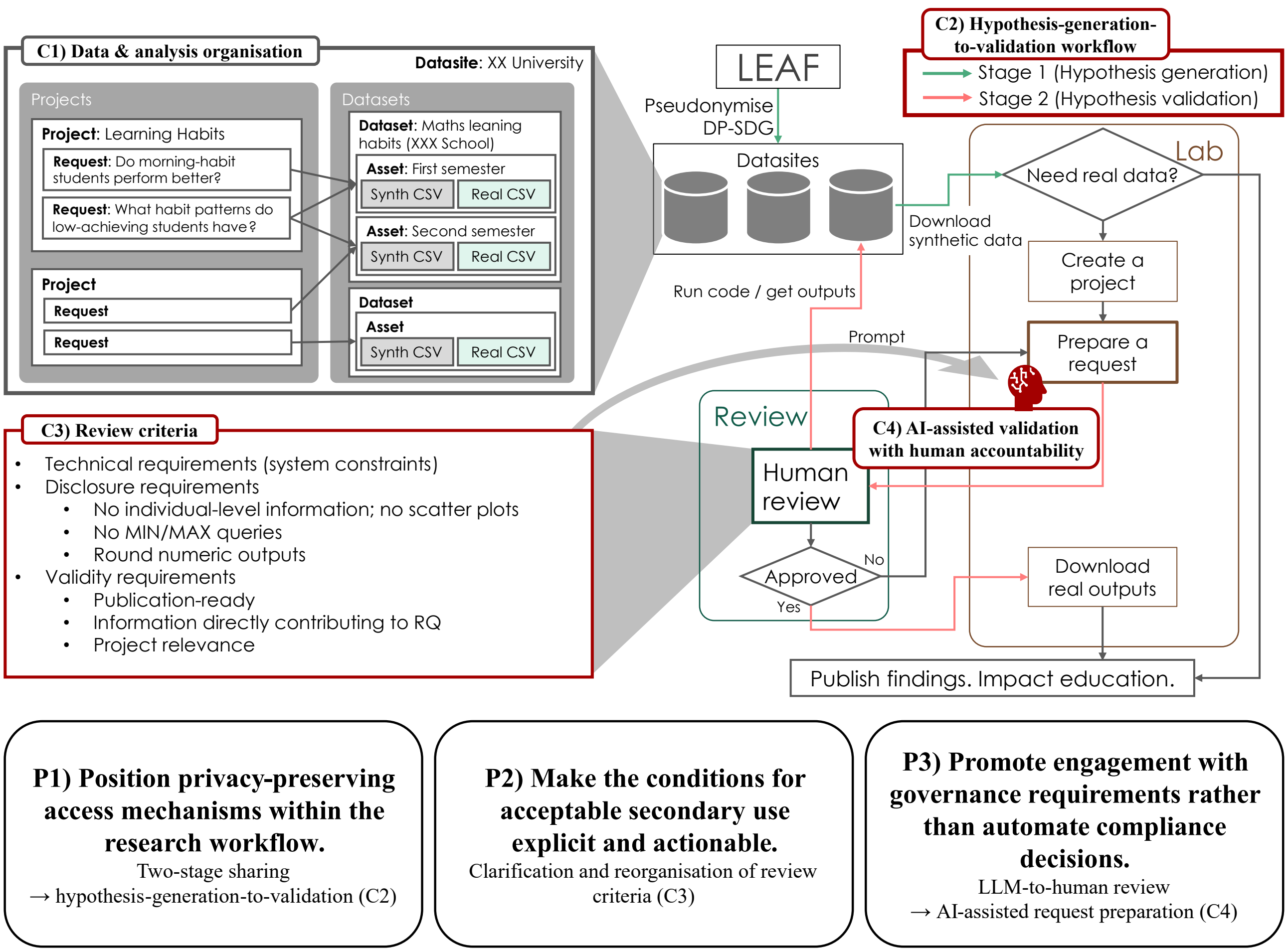

**Figure 11.** Empirically refined ReLEAF framework and design principles

## 5.2. Implications

The formative evidence indicates the feasibility of operationalising the two-stage mechanism through ReLEAF, while identifying remaining socio-technical challenges. ReLEAF remains dependent on expert review, and semantic complexity may increase review cost, potentially hindering scalability. Sustainable division of human labour and automation remains unresolved, and reviewer training is unaddressed in this work. Additionally, researchers still showed difficulty assessing disclosure risks. While P2 recommends making the review criteria explicit and actionable, reliable communication with researchers about these risks and requirements remains a challenge.

For practitioners, ReLEAF provides a usable framework to share ERWD that remains inaccessible in enclaves. Implementation requires two key considerations: 1) ReLEAF assumes the availability of an expert in statistical disclosure control, and 2) ReLEAF does not automatically bridge data custodians to data subjects (e.g. students and teachers). Importantly, it does not automatically eliminate the need for informed consent from data subjects. As evaluating disclosure risks requires some training and experience (Desai et al., 2016; Griffiths et al., 2024), this can also be a bottleneck for adoption. Studying the need for informed consent for sharing ERWD using ReLEAF is outside the scope of this work, and it should be determined in accordance with local data privacy regulations.

This work also provides broader implications for LA research in technical data-access mechanisms. As several participants mentioned, the discrepancy between real and synthetic data can heighten the need for costly real-data validation, damage users' trust in synthetic data and slow down the research process. Thus, research on improving DP-SDG techniques for ERWD is essential, which is particularly challenging due to the small-sample and high-dimensional problem (Heine et al., 2023; Ito et al., 2026c). In addition, although the findings suggest the potential of the AI assistant in developing educational-data analysis and engaging with data-access requirements, the generic zero-shot implementation limits its accuracy in judging disclosure risks. Calibration of the AI assistant requires technical development and evaluation.

### 5.3. Limitations and future work

The generalisability of the three evaluation studies is mostly limited due to the small sample sizes. The reliance on one reviewer limits the exploration of the potential reviewer-level variation in review results. Due to regulatory restrictions, evaluation is also limited to the Japanese context with a limited number of professional researchers involved. Further evaluation would be needed through real-world implementation in the wider LA community.

The ReLEAF framework should also be augmented in several directions. First, this work only considered single-datasite data sharing. Cross-datasite analytics require both technical augmentation such as federated learning and socio-technical evaluation of how that enhances secondary use of ERWD. Second, the DP-SDG process still relies on manual work of privacy experts and automating it can substantially reduce operational costs because ERWD is routinely collected. Third, sharing ERWD across platforms, including non-LEAF systems, is also an important direction for future work, aligning with the Open LA initiative (Muslim et al., 2020). Fourth, computationally assessing privacy risks of synthetic data and validation outputs is deferred to future work, especially for continually sharing longitudinal data. Finally, managing privacy budgets for DP is also an unaddressed governance question, when multiple datasets containing the same cohort are continuously released (Khavkin & Toch, 2025; Küchler et al., 2025).

## 6. Conclusion

This work proposes ReLEAF, a socio-technical framework that bridges data custodians and semi-trusted researchers for trustworthy secondary use of ERWD. Through three DSR cycles, we iteratively refined and formatively evaluated the proposed framework, resulting in the four key components and three design principles supported by formative evidence. ReLEAF therefore contributes both a concrete operational framework and transferable design knowledge for semi-trusted data access. At the same time, challenges remain around human-review scalability, disclosure-risk communication, AI calibration, and broader institutional deployment. This work provides practitioners with a usable framework for accessing otherwise inaccessible ERWD and highlights the need for further research on the socio-technical operationalisation of ERWD sharing.

## Declaration of Conflicting Interest

The authors declared no potential conflicts of interest with respect to the research, authorship, and/or publication of this article.

## Funding

This work was supported by CSTI SIP Grant Number JPJ012347, JSPS KAKENHI Grant Number 23H00505, 25KJ1515 and the Cabinet Office, Government of Japan, under the BRIDGE initiative through the MEXT measure "Building Educational AI Models and Developing a Human-Agent Collaborative AI Ecosystem for Educational Support."